\documentclass{article}
\usepackage{microtype}
\usepackage{graphicx}
\usepackage{booktabs}
\usepackage{multirow}
\usepackage{xcolor,array,amsmath,amssymb,amsfonts,mathtools}
\usepackage[acronym]{glossaries}

\newcommand{\Ecls}{\textcolor[HTML]{2B9348}{\textbf{\scriptsize[E]}}}
\newcommand{\Dcls}{\textcolor[HTML]{1F4E79}{\textbf{\scriptsize[D]}}}
\newcommand{\Pcls}{\textcolor[HTML]{E76F51}{\textbf{\scriptsize[P]}}}
\newcommand{\Hcls}{\textcolor[HTML]{6A4C93}{\textbf{\scriptsize[H]}}}
\newcommand{\tpot}{\mathrm{TPOT}}
\newcommand{\ttft}{\mathrm{TTFT}}
\newcolumntype{L}[1]{>{\raggedright\arraybackslash}p{#1}}

\usepackage{hyperref}

\usepackage[accepted]{mlsys2025_arxiv} 
\makeatletter
\renewcommand{\Notice@String}{Preprint. Under review at the Conference on Machine Learning and Systems (MLSys 2027)}
\makeatother

\usepackage{enumitem}

\setlist{
    nosep,
    topsep=2pt,
    partopsep=0pt,
    itemsep=1pt,
    parsep=0pt
}

\mlsystitlerunning{Evaluating Inference Compute for Generative AI: A Framework for Enterprise Workloads}

\makeglossaries

\begin{document}

\newacronym{Al-Jazeera}{Al-Jazeera}{Al-Jazeera Satellite Channel}
\newacronym{AraGPT-2}{AraGPT-2}{Arabic Generative Pre-trained Transformer 2}
\newacronym{AraGPT-3}{AraGPT-3}{Arabic Generative Pre-trained Transformer 3}
\newacronym{ArSAS}{ArSAS}{Arabic Speech-Act and Sentiment}
\newacronym{PATB}{PATB}{Penn Arabic Treebank}
\newacronym{BBC}{BBC}{British Broadcasting Corporation}
\newacronym{BERT}{BERT}{Bidirectional Encoder Representations from Transformers}
\newacronym{BPE}{BPE}{byte-pair encoding}
\newacronym{mBERT}{mBERT}{multi-lingual BERT}
\newacronym{RoBERTa}{RoBERTa}{A Robustly Optimized BERT Pre-training Approach}
\newacronym{CA}{CA}{Classical Arabic}
\newacronym{CC}{CC}{Common Crawl}
\newacronym{CC-100}{CC-100}{Common Crawl 100}
\newacronym{CNN}{CNN}{Cable News Network}
\newacronym{DA}{DA}{Dialectal Arabic}
\newacronym{EAPCOUNT}{EAPCOUNT}{English-Arabic Parallel Corpus of the United Nations Texts}
\newacronym{GenAI}{GenAI}{Generative AI}
\newacronym{GPT-1}{GPT-1}{Generative Pre-trained Transformer 1}
\newacronym{GPT-2}{GPT-2}{Generative Pre-trained Transformer 2}
\newacronym{GPT-3}{GPT-3}{Generative Pre-trained Transformer 3}
\newacronym{GPT}{GPT}{Generative Pre-trained Transformer}
\newacronym{GPU}{GPU}{Graphics Processing Unit}
\newacronym{HPL}{HPL}{High-Performance Linpack} 
\newacronym{IR}{IR}{Information Retrieval} 
\newacronym{KSUCCA}{KSUCCA}{King Saud University Corpus of Classical Arabic}
\newacronym{LM}{LM}{Language model}
\newacronym{LLM}{LLM}{Large Language Model}
\newacronym{LSTM}{LSTM}{Long-Short Term Memory}
\newacronym{ME}{ME}{Middle East}
\newacronym{MGB-2}{MGB-2}{Multi-Genre Broadcast-2}
\newacronym{MSA}{MSA}{Modern Standard Arabic}
\newacronym{MEGLN}{MEGLN}{MSA, Egyptian, Gulf, Levant \& North African}
\newacronym{MRC}{MRC}{Machine Reading Comprehension}
\newacronym{MT}{MT}{Machine Translation}
\newacronym{NN}{NN}{Neural Network}
\newacronym{NQ}{NQ}{Natural Questions}
\newacronym{NLP}{NLP}{Natural Language Processing}
\newacronym{OpenITI}{OpenITI}{Open Islamicate Texts Initiative}
\newacronym{OSAC}{OSAC}{Open-Source Arabic Corpora}
\newacronym{OSCAR}{OSCAR}{Open Super-large Crawled ALMAnaCH coRpus}
\newacronym{OSIAN}{OSIAN}{Open Source International Arabic News}
\newacronym{POS}{POS}{Part-of-speech}
\newacronym{PTB}{PTB}{Penn Tree Bank}
\newacronym{QA}{QA}{Question Answering} 
\newacronym{RNN}{RNN}{Recurrent Neural Network}
\newacronym{RT}{RT}{Russia Today}
\newacronym{SM}{SM}{Sentence Match}
\newacronym{SQuAD}{SQuAD}{Stanford Question Answering Dataset}
\newacronym{TPU}{TPU}{Tensor Processing Unit}
\newacronym{TFRC}{TFRC}{TensorFlow Research Cloud}
\newacronym{KSU}{KSU}{King Saud University}
\newacronym{WER}{WER}{Word Error Rate}
\newacronym{XNLI}{XNLI}{Cross-lingual Natural Language Inference}
\newacronym{HARD}{HARD}{Hotel Arabic Reviews Dataset}

\twocolumn[
\mlsystitle{Evaluating Inference Compute for Generative AI: A Framework for Enterprise Workloads}

\mlsyssetsymbol{equal}{*}

\begin{mlsysauthorlist}
\mlsysauthor{Abbas Raza Ali}{equal,abbas}
\mlsysauthor{Muhammad Ajmal Siddiqui}{ajmal}
\mlsysauthor{Moona Zahid}{moona}
\end{mlsysauthorlist}

\mlsysaffiliation{abbas}{Citigroup Inc., London, United Kingdom}
\mlsysaffiliation{ajmal}{Ernst \& Young LLP, London, United Kingdom}
\mlsysaffiliation{moona}{NVIDIA Corporation, London, United Kingdom}

\mlsyscorrespondingauthor{Abbas Raza Ali}{abbas.raza.ali@gmail.com}

\mlsyskeywords{GPU, Inference, Generative AI}

\vskip 0.3in

\begin{abstract}
LLM deployment is shifting from single-turn completion to \emph{agentic} trajectories in which a model plans, calls tools, reads results and reasons at test time before acting. This inverts the economics of inference hardware: chat serving amortises weight reads across large batches, whereas agent trajectories are sequentially dependent, run at effective batch one, and make per-token decode latency (TPOT) the dominant term in task completion time. Using a roofline analysis and a closed-form episode-latency model, we show why this regime favours accelerators that keep weights in on-die SRAM (Cerebras WSE-3/3T, Groq/NVIDIA LPU) or compiler-managed tiered memory (SambaNova SN40L/SN50), and why three vendor ecosystems converged in 2026 on \emph{disaggregated} prefill/decode serving. We show that per-step reliability compounds exponentially in trajectory length-a 2\% per-step failure rate erases a $2\times$ decode advantage for a 20-step agent-so determinism and tail latency are first-order performance variables. We then propose a four-layer evaluation framework (silicon, serving system, agent episode, enterprise) with a metric set built on goodput at an agentic SLO and cost per successful episode, a six-axis benchmark protocol over six task families, a paired-bootstrap statistical design, an attestation protocol for vendor-run benchmarks, and TCO, availability and adoption-timing models with explicit break-even conditions. All performance figures are public and labelled by evidence class; we state seven falsifiable hypotheses and the experiments that test them, and argue that the most likely original result is that token-throughput rankings diverge from cost-per-successful-task rankings on long-horizon work.
\end{abstract}
]

\printAffiliationsAndNotice{}

\section{Introduction}
Inference has become the dominant cost of deployed AI \Dcls: Cerebras's 2026 registration statement discloses a 750\,MW inference agreement with OpenAI \citep{cerebras2026s1}, and NVIDIA paid US\$20\,B for a non-exclusive licence to Groq's inference-only architecture \citep{groq2025,nvidia2026}. Neither transaction concerns training.

A second shift has changed \emph{what} is inferred. Chat completion is one forward pass followed by a token stream; a serving system batches hundreds of independent users to amortise reading model weights \citep{yu2022orca,kwon2023}. Agentic systems are different in kind \Ecls\ \citep{yao2023react,wang2024survey,yehudai2025}. A coding agent \citep{jimenez2024swebench} or a customer-operations agent \citep{yao2024tau} executes a \emph{trajectory}: it reasons-often emitting thousands of hidden tokens \citep{snell2024,deepseek2025,openai2024o1}-emits a structured tool call, waits, appends the result to its context and repeats for tens to thousands of steps \citep{kapoor2024,xu2024agentcompany}. Each step depends on the previous one; the effective batch of a trajectory is one; latency compounds; the KV cache grows monotonically.

These properties collide with how GPU inference has been made efficient. Continuous batching \citep{yu2022orca}, paged KV memory \citep{kwon2023} and chunked prefill \citep{agrawal2024sarathi} raise throughput by increasing concurrent sequences per weight read; they do nothing for an individual sequential trajectory and in the limit trade it away. The metric governing agent task time-single-stream decode latency-is precisely where HBM GPUs are weakest relative to SRAM-resident accelerators. The industry has moved decisively toward this reading \Dcls: Cerebras runs CS-3 as a \emph{decode} tier behind AWS Trainium3 \emph{prefill} \citep{aws2026}; NVIDIA's first Groq-derived product, the Groq\,3 LPX rack, is positioned as a decode co-processor for Vera Rubin \citep{nvidia2026}; SambaNova launched SN50 ``for agentic inference'' with an Intel Xeon\,6 heterogeneous design and was selected by JPMorganChase for on-premises inference \citep{sambanova2026}. Disaggregated, phase-specialised serving went from OSDI/ISCA prototype \citep{zhong2024distserve,patel2024splitwise} to the default architecture of three ecosystems within eighteen months.

Existing benchmarks measure silicon (roofline, peak specifications), serving systems (MLPerf, API probes) or agents (SWE-bench, $\tau$-bench) in isolation; enterprise decisions need all three plus the layer at which money and regulatory risk are accounted. This paper contributes:
\begin{enumerate}
\item a \textbf{workload characterisation} of agentic inference (taxonomy A1--A5) with a closed-form episode-latency model and a \textbf{failure-amplification model} showing that reliability compounds exponentially in trajectory length;
\item an \textbf{architectural taxonomy} of 2026 inference accelerators on two axes (weight residency, execution model) that explains the convergence on disaggregation (\S\ref{sec:arch});
\item a \textbf{four-layer evaluation framework}: layer-specific metrics, six task families $\times$ six sensitivity axes, a paired-bootstrap statistical design, and an attestation protocol for vendor-run benchmarks;
\item \textbf{TCO, availability and adoption-timing models} yielding explicit break-even conditions for high-capital single-appliance decode hardware in regulated enterprises (\S\ref{sec:enterprise});
\item \textbf{seven falsifiable hypotheses} with the experiments that test them (\S\ref{sec:hyp}).
\end{enumerate}
We tag claims by evidence class: \Ecls\ established research, \Dcls\ documented industry evidence (sub-classified as published specification, independent measurement, or vendor-reported), \Pcls\ proposed here, \Hcls\ hypothesis. We report no original measurements; \S\ref{sec:hyp} specifies the experiments that would make this an empirical paper.

\section{Background}\label{sec:bg}
\begin{figure*}[t]\centering
\includegraphics[width=\textwidth]{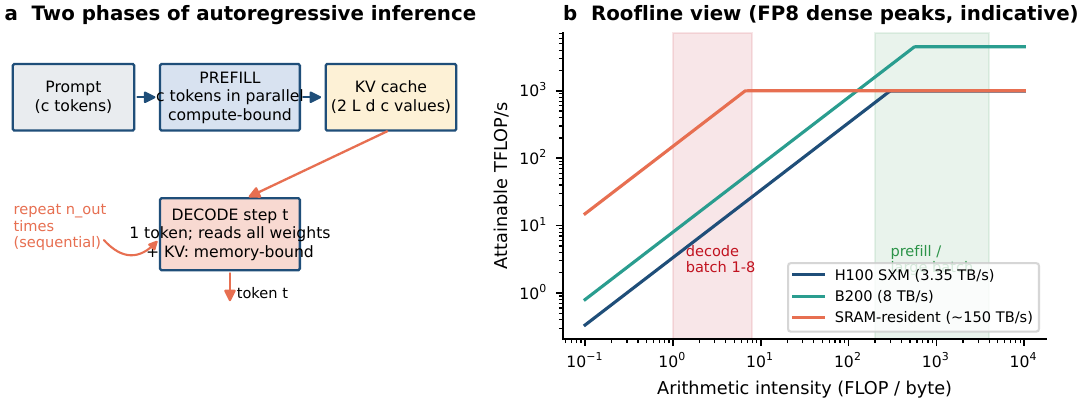}
\caption{\textbf{Prefill and decode have opposite bottlenecks.} (a) Prefill is compute-bound; decode re-reads all weights per token and is memory-bound. (b) Roofline \citep{williams2009} for three memory systems (FP8 dense peaks, indicative): at low-batch decode intensities (red) attainable throughput is set entirely by bandwidth; peak compute matters only at prefill/large-batch intensities (green). Device parameters are vendor specifications \Dcls.}\label{fig:phases}
\end{figure*}

\paragraph{Two phases, opposite bottlenecks \Ecls.}
Autoregressive Transformer inference \citep{vaswani2017} has two phases (Fig.~\ref{fig:phases}a). \emph{Prefill} processes all $c$ prompt tokens in parallel, reusing each weight $c$ times; it is compute-bound. \emph{Decode} produces one token per step and must re-read all $W$ bytes of weights plus the KV cache to perform a matrix--vector product; with batch $B$ each weight byte yields only $2B$ FLOPs, so decode is bandwidth-bound unless $B$ is large \citep{pope2023,gholami2024}. The per-token latency is bounded by
\begin{equation}\label{eq:tpot}
\tpot \;\ge\; \max\!\left\{\frac{W+\mathrm{KV}(c)}{BW},\; \frac{2NB}{F_{\mathrm{peak}}}\right\},
\end{equation}
where $N$ is the parameter count, $F_{\mathrm{peak}}$ peak arithmetic throughput and $BW$ weight-tier bandwidth. For a 70B FP8 model on one H100 (3.35\,TB/s) the first term is $\approx 21$\,ms, i.e.\ $\le 48$ tokens/s per stream before any tensor-parallel interconnect cost. On a device holding the same weights in on-die SRAM at $10^{2}$--$10^{4}$\,TB/s the bound is two to three orders of magnitude smaller \Dcls. The KV cache grows linearly: for $L$ layers, $n_{kv}$ KV heads \citep{ainslie2023gqa}, head dimension $d_h$ and $s$ bytes per element,
\begin{equation}\label{eq:kv}
\mathrm{KV}(c) = 2\,L\,n_{kv}\,d_h\,s\,c ,
\end{equation}
$\approx 160$\,kB/token for Llama-3.1-70B in FP8, so a 128k-token agent context occupies $\approx 20$\,GB-larger than any single chip's SRAM. IO-aware attention kernels \citep{dao2022flash,dao2024flash2} address the KV term but not the weight term.

\paragraph{Roofline and the memory wall \Ecls.}
On a roofline \citep{williams2009} (Fig.~\ref{fig:phases}b), batch-1--8 decode sits at 1--8 FLOP/byte, far left of any HBM device's ridge point ($\approx 300$ FLOP/byte for H100 at FP8). The memory wall \citep{wulf1995,gholami2024}-peak FLOPs growing $\sim3\times$ faster than DRAM bandwidth over two decades-is therefore the first-order determinant of single-stream decode speed. Three responses exist: raise $B$ (batching, at a latency cost), raise $BW$ (HBM3e$\to$HBM4, or on-die SRAM), or reduce $W$ per step (quantisation, MoE sparsity \citep{fedus2022}, speculative decoding \citep{leviathan2023}). GPUs pursue all three; SRAM accelerators bet on $BW$.

\paragraph{Batching, Little's law and goodput \Ecls.}
Little's law \citep{little1961} ties in-flight requests $N=\lambda W$ to arrival rate and residence time: higher $\lambda$ forces larger batches and, via Eq.~\ref{eq:tpot} and queueing, higher per-token latency. The latency--throughput frontier (Fig.~\ref{fig:frontier}) is intrinsic. DistServe and Splitwise \citep{zhong2024distserve,patel2024splitwise} introduced \emph{goodput}, the maximum request rate at which a stated fraction meets both a TTFT and a TPOT SLO:
\begin{equation}\label{eq:goodput}
\begin{split}G(\theta_1,\theta_2)=\max\lambda \;\text{ s.t. }\; &P(\ttft\le\theta_1)\ge 0.99,\\ &P(\tpot\le\theta_2)\ge 0.99 .\end{split}
\end{equation}
MLPerf Inference added an \emph{Interactive} scenario with tighter TTFT/TPOT limits in v5.0 and reports it separately from Offline and Server in v6.0 \citep{reddi2020mlperf,mlcommons2026}; a submission that leads Offline can be irrelevant to interactive deployment.

\paragraph{Disaggregated serving \Ecls$\to$\Dcls.}
Because the phases have opposite bottlenecks, co-locating them wastes one resource at any instant. Separating them onto different pools with KV transfer raised goodput $2$--$7\times$ at fixed SLOs \citep{zhong2024distserve,patel2024splitwise}. The 2026 vendor architectures (Fig.~\ref{fig:disagg}) are the industrial instantiation; we treat disaggregation as the reference architecture against which single-vendor appliances are evaluated.

\section{Characterising Inference}\label{sec:characterizing}

\paragraph{Episode-latency model \Pcls.}
Let an episode have $K$ sequential steps; at step $i$ the context is $c_i$, the model emits $n_i$ tokens (visible plus hidden), the environment takes $\tau_i$ to execute the tool call, $q_i$ is queueing delay and $\omega_i$ orchestration overhead (serialisation, schema validation, KV hand-off). Then
\begin{equation}\label{eq:episode}
\begin{split}T_{\mathrm{ep}}&=\sum_{i=1}^{K}\big[\ttft(c_i)+n_i\,\tpot(c_i)+\tau_i+q_i+\omega_i\big],\\ c_{i+1}&=c_i+n_i+o_i ,\end{split}
\end{equation}
with $o_i$ the tool-result length and $\ttft(c)\approx t_0 + c/R_{\mathrm{prefill}}$ (prefix caching reduces $c$ to the uncached suffix). Three consequences: (i) since $n_i$ multiplies TPOT and $K$ multiplies everything, decode dominates for any reasoning agent ($n_i\gtrsim100$); (ii) the benefit saturates when $n_i\,\tpot<\tau_i$-the tool-latency floor is the agentic Amdahl fraction; (iii) because context grows, capacity-tier bandwidth, not only weight bandwidth, sets late-episode TPOT.

\begin{figure*}[t]\centering
\includegraphics[width=0.92\textwidth]{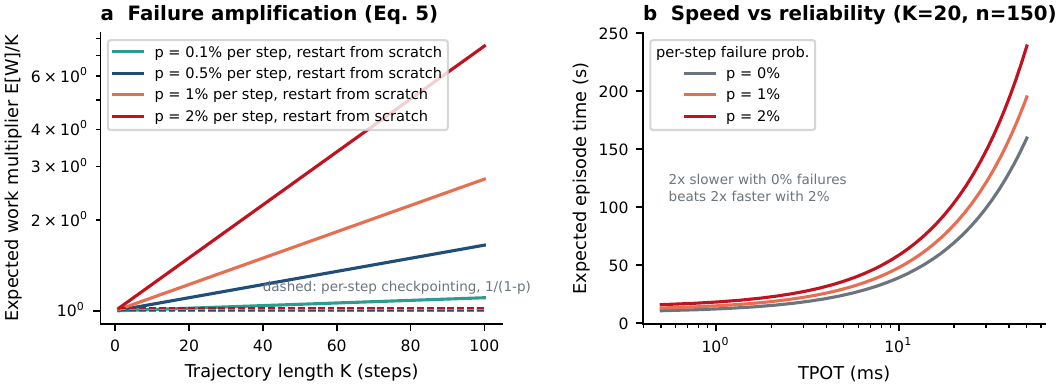}
\caption{\textbf{Reliability compounds along a trajectory \Pcls.} (a) Expected work multiplier (Eq.~\ref{eq:fail}) versus trajectory length for $p{=}0.1$--$2\%$, with and without per-step checkpointing. (b) Expected episode time for a 20-step agent versus TPOT and $p$: a faster, less reliable engine can lose to a slower, reliable one. Illustrative.}\label{fig:fail}
\end{figure*}

\paragraph{Reliability compounds \Pcls.}
Suppose each step fails independently with probability $p$ (timeout, malformed call, transient fault, non-deterministic divergence rejected by a verifier). If failure forces restart from scratch, an episode completes on one attempt with probability $(1-p)^K$; with per-step checkpointing only the failed step repeats:
\begin{equation}\label{eq:fail}
\frac{\mathbb{E}[W]}{K}=\begin{cases}(1-p)^{-K} & \text{restart from scratch},\\ (1-p)^{-1} & \text{per-step checkpointing}.\end{cases}
\end{equation}
At $p=1\%$ a 50-step agent without checkpointing does 65\% more work than ideal; at $2\%$, $2.7\times$; a 100-step agent at 2\% does $7\times$ (Fig.~\ref{fig:fail}a). Fig.~\ref{fig:fail}b shows the iso-time consequence: a decode engine twice as fast but with 2\% per-step failures (an immature compiler path, a flaky KV hand-off) is \emph{slower in expectation} for a 20-step agent than a half-speed reliable one. Hence tail latency and failure rate belong beside TPOT in Layer-1/2 measurement, and the \emph{determinism} of statically scheduled designs \citep{abts2020tsp} is a performance claim whose value scales with $K$ (hypothesis H4).

\section{Architectural Landscape, 2026}\label{sec:arch}
\begin{figure}[t]\centering
\includegraphics[width=\columnwidth]{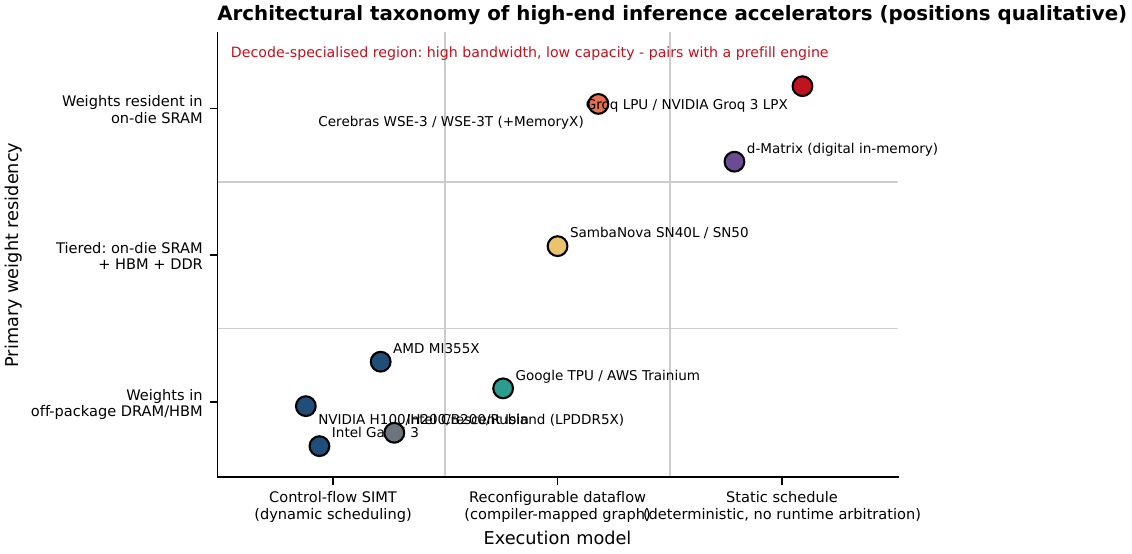}
\caption{\textbf{Architectural taxonomy \Pcls} (positions qualitative). The upper-right region is decode-specialised: extreme bandwidth, limited capacity, near-full utilisation at batch 1-a natural partner for an HBM-rich prefill engine rather than a replacement.}\label{fig:taxonomy}
\end{figure}

\begin{figure}[t]\centering
\includegraphics[width=\columnwidth]{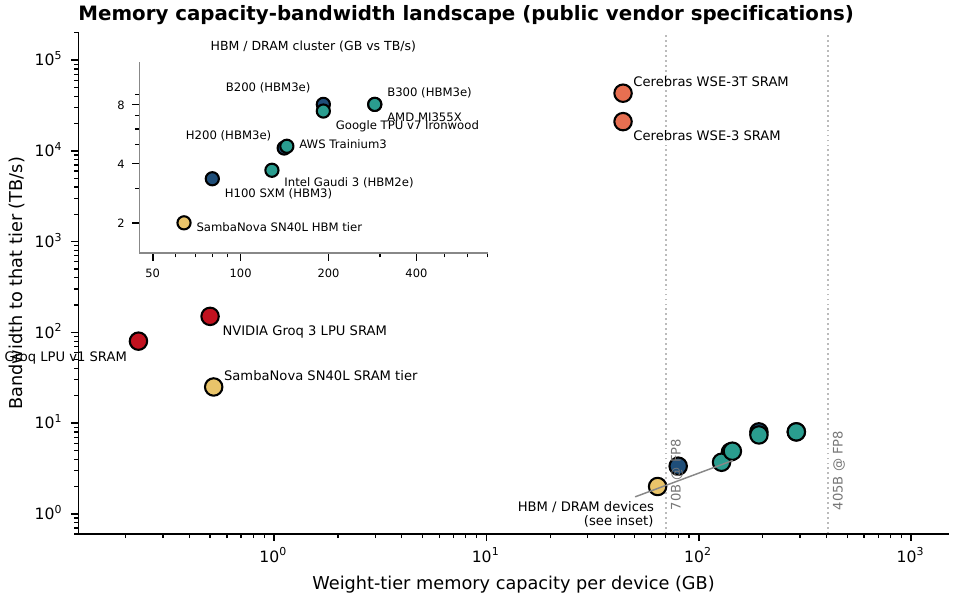}
\caption{\textbf{Capacity--bandwidth landscape \Dcls} (public specifications; log--log). No SRAM device holds a 70B FP8 model on one chip except the wafer-scale WSE-3; others shard across hundreds of chips or pair with a capacity tier. The 2--3 orders of bandwidth separating the clusters is the source of the decode-speed gap in Fig.~\ref{fig:speeds}.}\label{fig:mem}
\end{figure}

\paragraph{Two axes \Pcls.}
The domain-specific era \citep{hennessy2019} and the hardware lottery \citep{hooker2021} are both visible in Fig.~\ref{fig:taxonomy}, which places accelerators on two axes: \textbf{weight residency} (off-package HBM/DRAM; compiler-managed SRAM+HBM+DDR tiers; on-die SRAM), which sets $BW$ in Eq.~\ref{eq:tpot}; and \textbf{execution model} (dynamic SIMT; compiler-mapped spatial dataflow; fully static deterministic schedule \citep{abts2020tsp,abts2022}), which sets flexibility, batch-1 utilisation and latency variance. Table~\ref{tab:devices} gives public specifications; Fig.~\ref{fig:mem} plots the capacity--bandwidth landscape.

\begin{table*}[t]\centering\scriptsize
\caption{Public specifications of 2026 inference accelerators \Dcls\ (vendor-stated; SRAM bandwidths are aggregate on-die and not directly comparable to HBM). Sources: \citep{nvidia2026,nvidia2026nvl72,groq2025,intel2025,cerebras2024,cerebras2026cs4,prabhakar2024sn40l,sambanova2026,aws2026,mlcommons2026}.}\label{tab:devices}
\begin{tabular}{L{2.9cm}L{3.3cm}L{2.3cm}L{2.4cm}L{1.9cm}L{2.4cm}}\toprule
\textbf{Device} & \textbf{Weight tier / capacity} & \textbf{Bandwidth} & \textbf{Peak (dense, stated)} & \textbf{Execution} & \textbf{Status}\\\midrule
NVIDIA H100 / H200 & HBM3/3e, 80 / 141\,GB & 3.35 / 4.8\,TB/s & $\approx$1 PFLOPS FP8 & SIMT & Installed base\\
NVIDIA B200 / B300 & HBM3e, 192 / 288\,GB & $\approx$8\,TB/s & $\approx$4.5 PFLOPS FP8, FP4 & SIMT & Volume\\
NVIDIA GB200 NVL72 rack & 72 GPUs; 13.4\,TB HBM3e & 130\,TB/s NVLink aggregate & rack-scale & single NVLink domain & $\approx$120--140\,kW liquid rack\\
NVIDIA Rubin & HBM4, 288\,GB & $\approx$20+\,TB/s & $\sim$50 PFLOPS FP4 (claim) & SIMT & Ramp H2 2026\\
Groq\,3 LPU / LPX rack (NVIDIA) & on-die SRAM 0.5\,GB/LPU; 128\,GB/256-LPU rack & $\approx$150\,TB/s per LPU & n/d & static TSP & Decode co-processor, Q3 2026\\
Groq LPU v1 (GroqCloud) & on-die SRAM 230\,MB & 80\,TB/s & 188 TFLOPS FP16 & static TSP & Cloud only\\
Cerebras WSE-3 / CS-3 & on-die SRAM 44\,GB (+MemoryX) & 21\,PB/s & 125 PFLOPS FP16 (sparse) & spatial dataflow & Public; AWS decode tier\\
Cerebras WSE-3T / CS-4 (3 wafers) & 3$\times$44\,GB per rack & 43.2\,PB/s per wafer & 250 PFLOPS/wafer & dataflow; 2\,\textmu s wafer link & Announced Aug 2026\\
SambaNova SN40L & 520\,MB SRAM + 64\,GB HBM + 1.5\,TB DDR per socket; 16/rack & HBM $\approx$2\,TB/s & 638 TFLOPS BF16 & CGRA dataflow & Shipping; JPMC on-prem\\
SambaNova SN50 & same 3-tier; 256 sockets & n/d & 5$\times$ SN40L; FP8 & CGRA dataflow & H2 2026\\
Intel Gaudi 3 / Crescent Island & HBM2e 128\,GB / LPDDR5X 160\,GB & 3.7\,TB/s / low & 1.8 PFLOPS FP8 / n/d & tensor cores / Xe3P & Available (de-emphasised) / sampling\\
AMD MI355X & HBM3e 288\,GB & 8\,TB/s & $\approx$5 PFLOPS FP8 & SIMT & Leads MLPerf v6.0 Llama-2-70B\\
AWS Trainium3; Google TPU v7 & HBM3e 144\,GB; HBM 192\,GB & $\approx$4.9; 7.4\,TB/s & 2.5; 4.6 PFLOPS FP8 & systolic & Prefill partner to CS-3; GCP\\
\bottomrule\end{tabular}
\end{table*}

\paragraph{How each design pays for bandwidth \Dcls.}
\emph{Wafer-scale spatial dataflow} (Cerebras): a 46,225\,mm$^2$ wafer with $\approx$900k cores, 44\,GB SRAM and 21\,PB/s on a 2-D mesh \citep{lie2023,cerebras2024}; models beyond SRAM stream from MemoryX. Costs: a single fault domain, a 23--25\,kW liquid-cooled chassis and a compiler that maps each model onto the fabric. CS-4 (Aug 2026) runs the same 5\,nm silicon at a higher operating point-power delivery 0.5\,mm from the wafer, redesigned cooling-to double stated compute and bandwidth per wafer, with three wafers per rack at 2\,\textmu s wafer-to-wafer latency; the vendor reports $>$4,400 tok/s/user on GPT-OSS-120B and announces prefill partnerships with AWS Trainium and AMD Helios \citep{cerebras2026cs4}. Operating as a decode tier concedes that prefill on wafer-scale SRAM is not the economic optimum.
\emph{Static tensor streaming} (Groq$\to$NVIDIA LPU): a single VLIW-style core with a compile-time schedule, no caches and no arbitration, so every run takes the same cycles \citep{abts2020tsp,abts2022}; weights in SRAM (230\,MB v1; 500\,MB Groq\,3) mean a 70B model spans hundreds of chips; batch-1 utilisation is near-complete, which is why the architecture led single-stream leaderboards \citep{artificialanalysis}. Limits are capacity (Eq.~\ref{eq:kv}) and prefill; NVIDIA ships Groq\,3 strictly as a decode co-processor, with later generations adopting NVLink and NVFP4 \citep{nvidia2026}.
\emph{Tiered dataflow} (SambaNova): a CGRA with 520\,MB scratchpad SRAM, 64\,GB HBM and up to 1.5\,TB DDR per socket \citep{prabhakar2024sn40l}; spatial operator placement keeps activations on-chip while the DDR tier makes many models-or a trillion-parameter MoE-resident and hot-swappable (``composition of experts''), which maps directly onto multi-agent topologies. SN50 keeps the hierarchy, adds FP8, quintuples throughput and scales the shared memory space to 256 sockets \citep{sambanova2026}.
\emph{HBM GPUs and hyperscaler ASICs}: the general-purpose baseline with the richest ecosystem and the only platform on which training, fine-tuning, multimodal and inference coexist. Rack-scale NVLink domains (GB200/GB300 NVL72: 72 GPUs, 13.4\,TB HBM3e, 130\,TB/s aggregate NVLink \citep{nvidia2026nvl72}) deserve treatment as a distinct class: they reduce the interconnect bound that Pope et al.\ \citep{pope2023} identify and shift the break-even of \S\ref{sec:enterprise} for very large models. AMD MI355X and TPU v7 / Trainium3 \citep{jouppi2023tpuv4} occupy the same cell. The GPU vendors are nonetheless adding SRAM decode tiers rather than closing the single-stream gap with HBM alone.

\begin{figure}[t]\centering
\includegraphics[width=\columnwidth]{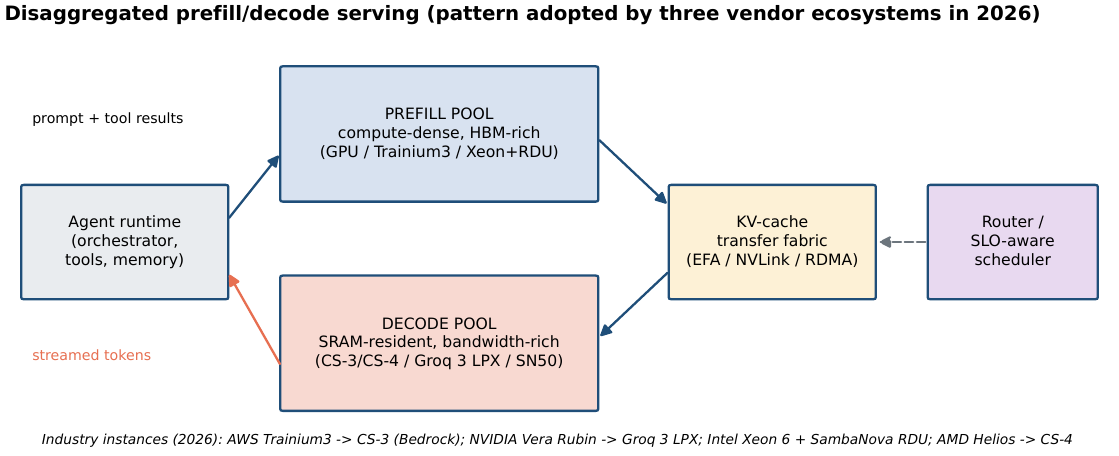}
\caption{\textbf{Disaggregated prefill/decode serving}: established in research \citep{zhong2024distserve,patel2024splitwise}; adopted in 2026 by AWS+Cerebras \citep{aws2026}, NVIDIA Rubin+Groq\,3 LPX \citep{nvidia2026} and Intel+SambaNova \citep{sambanova2026}.}\label{fig:disagg}
\end{figure}

\paragraph{Convergence on disaggregation \Dcls.}
The three 2026 announcements share the structure of Fig.~\ref{fig:disagg}. AWS states $\sim5\times$ more high-speed token capacity in the same footprint \citep{aws2026}; NVIDIA claims $35\times$ decode throughput per megawatt for LPX over HBM GPUs \citep{nvidia2026}-vendor claims, untested by us. For an evaluator the unit of comparison is no longer a chip but a \emph{phase-specialised pool pair}, and the KV-transfer fabric is a component whose bandwidth and latency must be benchmarked (axis A5).

\section{Documented Evidence and Its Limits}\label{sec:evidence}
\begin{figure}[t]\centering
\includegraphics[width=\columnwidth]{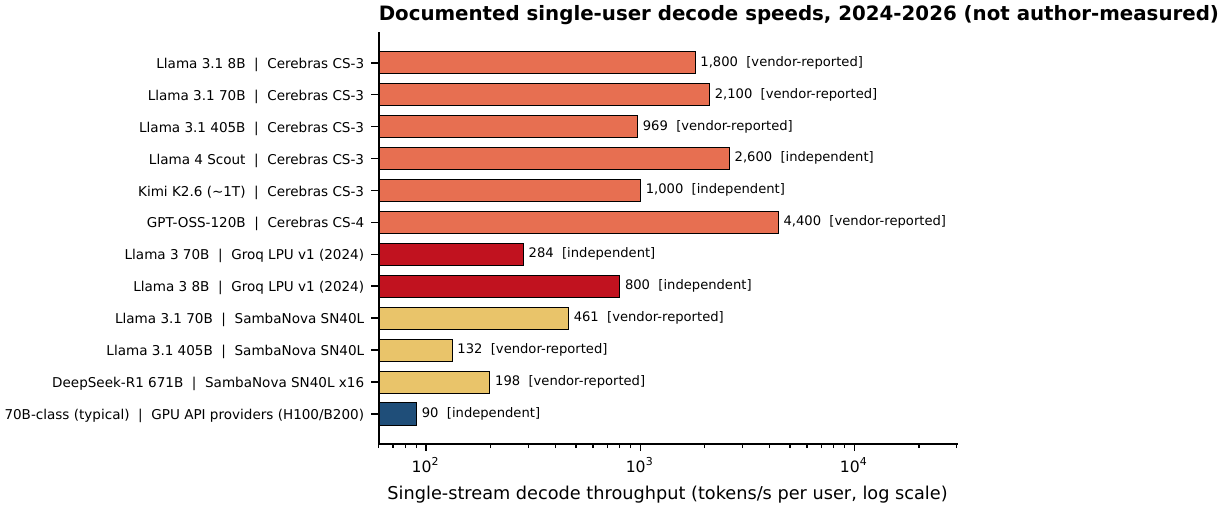}
\caption{\textbf{Documented single-user decode throughput, 2024--2026 \Dcls} (log scale; sources in Table~\ref{tab:speeds}). Not measured by the authors; not a common protocol.}\label{fig:speeds}
\end{figure}

Fig.~\ref{fig:speeds} and Table~\ref{tab:speeds} collect public single-user decode figures, labelled by sub-class.

\begin{table}[t]\centering\scriptsize
\caption{Documented decode-speed evidence \Dcls. D-indep: Artificial Analysis or MLCommons; D-vendor: press releases.}\label{tab:speeds}
\begin{tabular}{L{1.6cm}L{1.5cm}L{1.2cm}L{2.2cm}}\toprule
\textbf{Model} & \textbf{Hardware} & \textbf{tok/s/user} & \textbf{Source, class}\\\midrule
Llama 3.1 8B / 70B & Cerebras CS-3 & 1,800 / 2,100 & \citep{cerebras2024} D-vendor\\
Llama 3.1 405B & Cerebras CS-3 & 969 & \citep{cerebras2024} D-vendor\\
Llama 4 Scout & Cerebras CS-3 & $>$2,600 & \citep{artificialanalysis} D-indep\\
Kimi K2.6 ($\sim$1T) & Cerebras CS-3 & $\approx$1,000 & \citep{cerebras2026s1,artificialanalysis} D-indep\\
GPT-OSS-120B & Cerebras CS-4 & $>$4,400 & \citep{cerebras2026cs4} D-vendor\\
Llama 3 70B & Groq LPU v1 & 284 (TTFT 0.3\,s) & \citep{groq2024aa} D-indep\\
Llama 3 8B & Groq LPU v1 & $\approx$800 & \citep{artificialanalysis} D-indep\\
Llama 3.1 70B / 405B & SambaNova SN40L & 461 / 132 & \citep{sambanova2026} D-vendor\\
DeepSeek-R1 671B & SN40L $\times$16 & 198 & \citep{sambanova2026} D-vendor\\
70B-class & GPU API providers & $\approx$30--200 & \citep{artificialanalysis} D-indep\\
Llama 2 70B & AMD MI355X$\times$8 & aggregate only & \citep{mlcommons2026} D-indep\\
\bottomrule\end{tabular}
\end{table}

\begin{figure}[t]\centering
\includegraphics[width=\columnwidth]{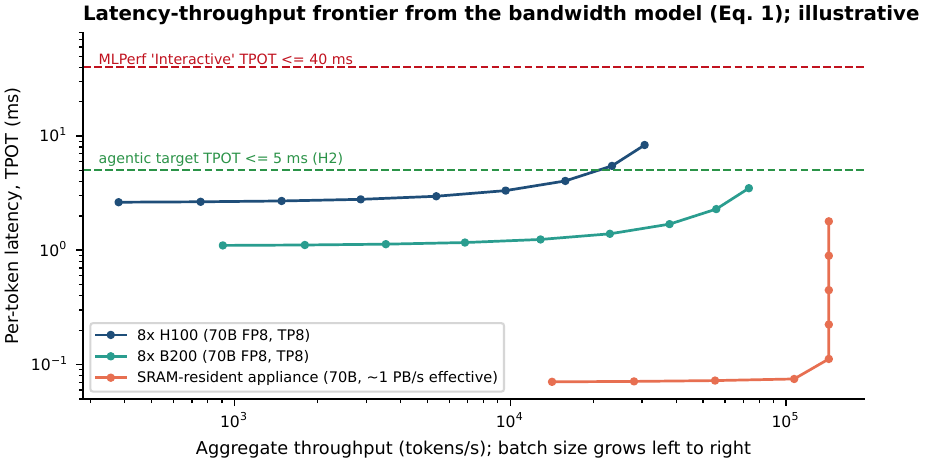}
\caption{\textbf{The latency--throughput frontier is intrinsic to batching \Ecls} (bandwidth model of Eq.~\ref{eq:tpot}, 70B FP8; illustrative). GPUs reach high aggregate throughput only by moving right, raising TPOT; an SRAM pool sits below the MLPerf Interactive limit \citep{mlcommons2026} at all batch sizes. A single-stream comparison that does not fix the operating point is under-specified.}\label{fig:frontier}
\end{figure}

These numbers come from at least four incompatible regimes: vendor lab runs with unstated batch and precision; public endpoints whose operators choose their own batching point; MLPerf submissions that fix model, accuracy and latency limits but report aggregate throughput \citep{reddi2020mlperf}; and comparisons against unspecified ``leading GPUs''. Two biases follow. Single-stream speed on an SRAM device is a \emph{hardware} property; on a GPU endpoint it is a \emph{policy} property (Fig.~\ref{fig:frontier})-an operator can always shrink batches-so a fair comparison must hold goodput or cost constant. And none of the figures measure Eq.~\ref{eq:episode} under realistic context growth and tool interleaving. Our protocol is designed to close both gaps.

\begin{table}[t]\centering\scriptsize
\caption{Metric set by layer (E established, P proposed).}\label{tab:metrics}
\begin{tabular}{L{0.6cm}L{2.4cm}L{3.5cm}c}\toprule
\textbf{Layer} & \textbf{Metric} & \textbf{Definition / why} & \\\midrule
L1 & weight-tier BW, capacity & Eq.~\ref{eq:tpot}--\ref{eq:kv}; STREAM-style kernel & E\\
L1 & tokens/J (batch 1, 8) & wall-plug energy \citep{luccioni2024,samsi2023} & E\\
L1 & determinism & CoV of TPOT over $10^3$ identical runs & P\\
L2 & TTFT, TPOT P50/P99 vs context & MLPerf Interactive definitions & E\\
L2 & goodput $G(\theta_1,\theta_2)$ & Eq.~\ref{eq:goodput} at the agent SLO & E\\
L2 & prefix-cache hit rate; KV-transfer P99 & context grows each step; disaggregation adds a hop & P\\
L2 & structured-decoding overhead & TPOT constrained / unconstrained & P\\
L2 & per-step failure $p$; recovery time & enters Eq.~\ref{eq:fail} with exponent $K$ & P\\
L2 & $\omega$, $q$ & wall-clock minus (TTFT $+ n\,$TPOT $+\tau$) & P\\
L3 & $T_{\mathrm{ep}}$ P50/P95 & Eq.~\ref{eq:episode} on a fixed harness & P\\
L3 & success $S$; tokens/successful task & \citep{kapoor2024,yao2024tau} & E\\
L3 & tool-interleave efficiency & decode idle fraction (Amdahl floor) & P\\
L4 & cost per successful episode & Eq.~\ref{eq:cost} & P\\
L4 & time-to-reproducible-evidence & months to in-house benchmark & P\\
L4 & availability-adjusted capacity & Eq.~\ref{eq:avail} & E\\
L4 & third-party-risk score, exit cost & Appendix Table~A2 instruments & D\\
\bottomrule\end{tabular}
\end{table}

\paragraph{Cost and energy identities \Pcls.}
With success rate $S$, mean tokens $T_{\mathrm{in}},T_{\mathrm{out}}$ per attempt at effective prices $p_{\mathrm{in}},p_{\mathrm{out}}$ and tool cost $C_{\mathrm{tool}}$,
\begin{equation}\label{eq:cost}
\begin{split}C_{\mathrm{ep}}&=\frac{T_{\mathrm{in}}p_{\mathrm{in}}+T_{\mathrm{out}}p_{\mathrm{out}}+C_{\mathrm{tool}}}{S},\\ \eta&=\frac{S}{E_{\mathrm{decode}}+E_{\mathrm{prefill}}+E_{\mathrm{tools}}}\end{split}
\end{equation}
($C_{\mathrm{ep}}$: cost per successful episode; $\eta$: successful episodes per kWh). A faster but less accurate configuration can raise cost, and hidden reasoning tokens are a first-order cost term.

\paragraph{Protocol: task families $\times$ sensitivity axes \Pcls.}
A protocol needs two orthogonal specifications-\emph{what} the agent does and \emph{which variable} is swept; conflating them is why most published comparisons cannot separate a hardware effect from an agent-policy effect. Table~\ref{tab:families} defines six task families spanning A1--A5 and Table~\ref{tab:axes} six axes, each varied at a stated baseline. A minimum credible study covers S1, S3, S4 against A1, A3, A5 (plus A6 for reliability claims). The protocol is executable against a cloud endpoint, a loaner appliance or a production pool; a cloud-agnostic enterprise \gls{GenAI} platform that exposes models through a single provider-independent gateway \citep{Abbas2024} is a convenient harness.

\begin{table}[t]\centering\scriptsize
\caption{Task families \Pcls\ (add enterprise-private sets to avoid over-fitting).}\label{tab:families}
\begin{tabular}{L{1.7cm}L{0.6cm}L{1.8cm}L{2.7cm}}\toprule
\textbf{Family} & \textbf{Tax.} & \textbf{Stress} & \textbf{Outputs; reference envs}\\\midrule
S1 single-step & A1 & 1k--128k in; 64--2k out & TTFT, TPOT, tok/J; MLPerf Interactive\\
S2 long-context RAG & A1--2 & 2k--128k retrieved & TTFT, KV footprint, accuracy \citep{liu2024lost}\\
S3 tool-augmented agent & A2 & 5--30 steps, 1--8 tools & $T_{\mathrm{ep}}$, $S$, tool share; $\tau$-bench, GAIA\\
S4 long-horizon agent & A5 & 20--200 steps, retries 0--3 & $S$, time-to-success, P99, $p$; SWE-bench, WebArena, TheAgentCompany\\
S5 concurrent agents & A2$\times$conc. & 1--256 trajectories & goodput, P95/P99, fairness\\
S6 multi-agent & A3 & 2--8 agents; shared vs private KV & completion, $\omega$, network bytes; \citep{wu2023autogen,cemri2025}\\
\bottomrule\end{tabular}
\end{table}

\begin{table}[t]\centering\scriptsize
\caption{Sensitivity axes \Pcls. Baseline: ReAct loop \citep{yao2023react} over a fixed tool set, run identically on every system.}\label{tab:axes}
\begin{tabular}{L{1.3cm}L{2.2cm}L{1.6cm}L{1.9cm}}\toprule
\textbf{Axis} & \textbf{Levels} & \textbf{Metrics} & \textbf{Hypothesis}\\\midrule
A1 context & 2k, 8k, 32k, 128k & TTFT, TPOT vs $c$ & H1\\
A2 output length & 16, 128, 1k, 8k & TPOT stability, tok/J & static devices flat\\
A3 concurrency & 1--256 agents & $G$ at $\theta_2{=}10$, 40\,ms; P99 & H2\\
A4 format & text / JSON-schema / code & TPOT ratio, validity & overhead in runtime, not silicon\\
A5 tool interleave & $K{=}5,20,50$; $\tau{=}50$\,ms, 0.3\,s, 2\,s & $T_{\mathrm{ep}}$, idle fraction, KV-transfer P99 & H3\\
A6 faults \& reasoning budget & injected $p{=}0/0.5/1/2\%$; budget 1/2/4$\times$ & measured $p$, $\mathbb{E}[W]/K$, CoV, Eq.~\ref{eq:cost} & H4\\
\bottomrule\end{tabular}
\end{table}

\paragraph{Statistical design \Pcls.}
Cross-platform inference comparisons are routinely under-powered and confounded. We require: \textbf{paired, not pooled}-every task instance runs on every system with identical checkpoint hash, tokenizer, precision, sampling and agent policy; report paired mean/median differences with bootstrap 95\% intervals ($\ge$2,000 resamples) and a Wilcoxon statistic \citep{efron1993}; \textbf{two configurations per platform}-a \emph{capability-equivalent} configuration (identical model/precision/features everywhere) measures architecture, a \emph{system-optimised} one (each vendor's best path: FP4, speculation, fused kernels) measures the product; reporting only one measures software effort and calls it silicon; \textbf{factorial sweeps with interactions}-at minimum concurrency$\times$trajectory length, because a platform that wins short independent requests can collapse when many long-lived agents contend for KV capacity \citep{montgomery2019}; \textbf{tails and variance components}-P50/P95/P99, timeout and failure rates, run-to-run CoV, variance by task family, since aggregate leaderboards hide platform-by-task interactions \citep{kapoor2024,yehudai2025}; \textbf{warm-up, repetition and interleaving}-$\ge$30 repetitions per latency cell, $\ge$50--100 trajectories per family, systems interleaved in time to control endpoint load and software drift; and a \textbf{provenance record} (Appendix Table~A5) with every result.

\paragraph{Attestation of vendor-run benchmarks \Pcls.}
In regulated firms the first evidence is almost always produced off-premises by the vendor (\S\ref{sec:enterprise}); this is legitimate if \emph{attested}. Following model-risk practice \citep{sr117}, a result that cannot be reproduced from its documentation is an assertion. Required artefacts: (1) checkpoint hash, precision, quantisation and speculative configuration; (2) per-request timestamps at request, first token and every token, so the enterprise recomputes TTFT/TPOT/goodput itself; (3) batching and concurrency policy on the vendor system \emph{and} any baseline (fixing the point on Fig.~\ref{fig:frontier}); (4) the enterprise's own sanitised prompt/trajectory set; (5) PDU or wall power at $\ge$1\,Hz; (6) software versions and a statement of what differs on-premises; (7) independent witness for material procurements. ``$N\times$ faster than GPUs'' without these is marketing; with them it is recomputable evidence.

\paragraph{From measurement to decision \Pcls.}
We recommend a gated funnel: workload definition (trajectory traces, SLOs) $\to$ capability screen (model coverage, both configurations definable) $\to$ performance (A1/A3/A5, paired $T_{\mathrm{ep}}$ gain) $\to$ reliability (A6: $p$, CoV, P99) $\to$ economics (Eqs~\ref{eq:cost},\ref{eq:tco},\ref{eq:avail}; capacity ratio $>$ capital ratio at realistic utilisation) $\to$ governance (Appendix Table~A2; Eq.~\ref{eq:adopt}) $\to$ on-premises pilot. Cheap gates run against vendor endpoints first. Where a scalar is demanded, apply it only to gate survivors:
\begin{equation}\label{eq:score}
U=w_S S - w_T\tilde T_{\mathrm{ep}} - w_C\tilde C - w_E\tilde E - w_R\tilde R - w_D\tilde D ,
\end{equation}
with normalised episode time, cost, energy, reliability penalty (Eqs~\ref{eq:fail},\ref{eq:avail}) and deployment friction (Eq.~\ref{eq:adopt}), and published weights. The quantities are incommensurable; the lexicographic gates dominate when score and gates disagree.

\section{Enterprise Deployment and Economics}\label{sec:enterprise}
\paragraph{Time-to-reproducible-evidence \Dcls$\to$\Pcls.}
Financial institutions treat a new hardware vendor-especially one with a management plane, remote-support path or telemetry-as an ICT third-party arrangement subject to due diligence, contractual controls, concentration-risk assessment and exit planning \citep{eba2019}; the AI-specific instruments \citep{nist2024genai} add logging and robustness duties (mapped in Appendix Table~A2). Indicatively, vendor-run evidence is available within $\sim$3 months and reproducible in-house evidence after $\sim$9 (Appendix Fig.~A1). Let $u_c,u_i$ be the per-month value of challenger and incumbent once in production, $\tau_c,\tau_i$ their times to production and $T$ the horizon; ignoring discounting the challenger is justified only if
\begin{equation}\label{eq:adopt}
u_c\,(T-\tau_c) > u_i\,(T-\tau_i)\iff \frac{u_c}{u_i} > \frac{T-\tau_i}{T-\tau_c}.
\end{equation}
For $T{=}36$, $\tau_i{=}1$, $\tau_c{=}9$ months the challenger must deliver $\ge 1.3\times$ realised value to break even-more with discounting and onboarding-failure risk. Equation~\ref{eq:adopt} makes lead time a performance variable and explains why hyperscaler channels that reuse an existing third-party arrangement can be decisive even when not the fastest route to the silicon. Exit cost is bounded because HBM GPUs remain a universal fallback: any decode-tier workload runs, more slowly, on GPUs, so exit is latency regression plus re-tuning, priced via Eq.~\ref{eq:episode}.

\begin{figure}[t]\centering
\includegraphics[width=\columnwidth]{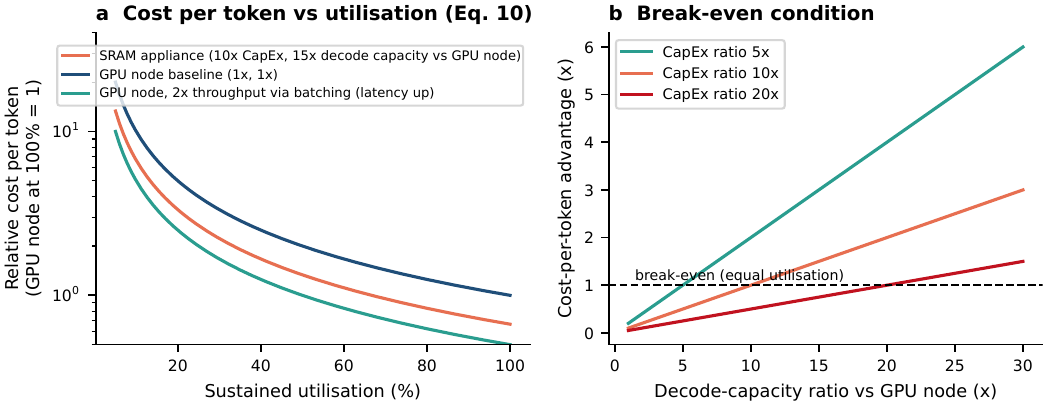}
\caption{\textbf{TCO model \Pcls} (Eq.~\ref{eq:tco}; illustrative). (a) Relative cost per token vs utilisation for a GPU node and an appliance with $10\times$ CapEx and $15\times$ decode goodput at the agentic SLO. (b) Break-even: advantage $=$ capacity ratio $/$ capital ratio. 2024 public price points ($\approx$US\$2--3\,M per CS-3 vs US\$0.3--0.6\,M per 8-GPU server) imply a capital ratio near $5$--$10\times$; the required capacity ratio is then a measurable threshold.}\label{fig:tco}
\end{figure}

\paragraph{TCO with explicit break-even \Pcls.}
Let CapEx be acquisition cost, $H$ accounting life in hours, $\mathrm{OpEx}_h$ hourly operating cost, $R$ sustainable decode goodput at the SLO and $U\in(0,1]$ utilisation:
\begin{equation}\label{eq:tco}
C_{\mathrm{tok}}=10^{6}\,\frac{\mathrm{CapEx}/H+\mathrm{OpEx}_h}{R\,U}.
\end{equation}
At equal utilisation and OpEx an appliance ($a$) beats a GPU node ($g$) iff $R_a/R_g>\mathrm{CapEx}_a/\mathrm{CapEx}_g$ (Fig.~\ref{fig:tco}b). Three points: $R$ must be measured at the agentic SLO, where GPUs cannot batch and the capacity ratio is largest; because CapEx$/H$ dominates for expensive single-purpose hardware, utilisation is decisive (Fig.~\ref{fig:tco}a)-an appliance at 30\% can cost more per token than GPUs at 80\% regardless of speed; and consumption pricing (Bedrock, GroqCloud, SambaCloud) converts CapEx$/H$ into a token price and removes utilisation risk at the cost of data-residency assessment. Power, not floor space, is the scarce facility resource \citep{iea2025,barroso2018}; liquid cooling no longer differentiates wafer-scale from Blackwell/Rubin racks (Appendix Table~A3).

\paragraph{Concentration risk \Ecls$\to$\Pcls.}
A 15U appliance replacing four GPU racks concentrates their failure exposure. With $A=\mathrm{MTBF}/(\mathrm{MTBF}+\mathrm{MTTR})$,
\begin{equation}\label{eq:avail}
A_{N+1}=1-(1-A)^2,\qquad \text{lost capacity-h/yr}\approx 8760\,(1-A)\,f ,
\end{equation}
where $f$ is the capacity fraction one failure removes (1 for a lone appliance; $1/N$ for an $N$-node pool). A lone appliance with 24\,h MTTR loses $\approx$24 capacity-hours per failure-year all at once; an 8-node pool loses one-eighth for the same period (Appendix Fig.~A2). The remedy-a second appliance-doubles CapEx in Eq.~\ref{eq:tco}. Wafer-scale designs tolerate \emph{manufacturing} defects with spare cores \citep{lie2023} but not \emph{field} failures of power, cooling or the single fabric; field MTBF data are not public \Hcls. A GPU fallback is thus the availability strategy that keeps the appliance economical, and operational-resilience regimes \citep{dora2022,pra2021} expect it documented.

\paragraph{Vendor landscape, Sept 2026 \Dcls.}
The stand-alone-startup risk of 2024 has largely resolved: Cerebras is public (Nasdaq: CBRS) with OpenAI and AWS as anchors, disclosing customer concentration, wafer-yield and export-licence risks \citep{cerebras2026s1}; Groq's team and IP licence sit inside NVIDIA, with GroqCloud continuing independently and the licence structure under Senate/FTC scrutiny \citep{groq2025}; SambaNova raised US\$350\,M and then US\$1\,B at US\$11\,B after Intel acquisition talks lapsed, with JPMorganChase as a named on-premises customer \citep{sambanova2026}; Intel de-emphasised Gaudi for Crescent Island and heterogeneous designs \citep{intel2025}. Every path to SRAM decode now also runs through a hyperscaler or incumbent channel, shortening the regulated onboarding path (Eq.~\ref{eq:adopt}) at the price of cloud residency assessments (Appendix Table~A4).

\section{Hypotheses and Threats to Validity}\label{sec:hyp}
Table~\ref{tab:hyp} states falsifiable predictions and their tests; Appendix Table~A6 gives the full workload$\times$architecture matrix as signed predictions of the paired difference.

\begin{table}[t]\centering\scriptsize
\caption{Hypotheses and the experiments that test them \Hcls.}\label{tab:hyp}
\begin{tabular}{L{0.4cm}L{3.7cm}L{2.4cm}}\toprule
& \textbf{Hypothesis} & \textbf{Test / falsifier}\\\midrule
H1 & On SRAM decode engines TPOT is flat in context until KV exceeds on-die capacity, then step-wise; on HBM GPUs it rises smoothly. & A1, batch 1 / smooth SRAM degradation\\
H2 & At $\theta_2\le10$\,ms P99, SRAM pools reach $\ge5\times$ the goodput per rack of Blackwell-class pools; at $\theta_2\ge100$\,ms $<2\times$. & A3 sweep / ratio invariant to $\theta_2$\\
H3 & Episode-time gains saturate once $n\,\tpot<\tau_{\mathrm{tool}}$; for $\tau\approx0.3$--$2$\,s the value of TPOT $<2$\,ms is negligible. & A5 / continued gains below 2\,ms\\
H4 & Deterministic execution yields lower $p$ and TPOT CoV, hence lower $\mathbb{E}[W]/K$, on S4/S6. & A6 fault injection / no difference\\
H5 & Disaggregated pairs give lower cost per successful episode than either homogeneous pool for A2--A4, not A1. & Eq.~\ref{eq:cost} across classes / homogeneous wins\\
H6 & In regulated firms time-to-reproducible-evidence, not peak throughput, predicts adoption (Eq.~\ref{eq:adopt}). & Retrospective procurement analysis\\
H7 & Token-throughput and cost-per-successful-task rankings diverge on S4 once $p$ and reasoning budget are included. & Spearman $\rho$ between rankings / $\rho\approx1$\\
\bottomrule\end{tabular}
\end{table}

H7 is the single result most likely to constitute an original systems contribution: it would show that trajectory-level evaluation changes conclusions that appear stable under token-level benchmarks. \textbf{Threats to validity}: no original measurements (all figures public and heterogeneous); vendor-stated specifications for products not yet shipping in volume (Groq\,3 LPX, SN50, Rubin, CS-4); Eq.~\ref{eq:episode} treats step counts and token lengths as exogenous, whereas faster inference changes agent behaviour (more sampling, more reflection); strict capability-equivalent configurations may not exist for every platform pair, which is why both configurations are reported; public agent environments encode assumptions that may not match enterprise workflows and an agent policy can over-fit a toolchain, so cross-policy validation is required; and measurement boundaries (device/PDU/wall power; API-path vs in-rack latency) are not interchangeable and must be stated. The minimum experiment that converts this framework into an empirical paper is small: three OpenAI-compatible endpoints (a GPU provider, an SRAM appliance cloud, and a disaggregated path), families S3--S4, axes A1/A3/A5/A6, paired design with bootstrap intervals, fitting Eq.~\ref{eq:episode} and testing H1--H3 and H7.


\section{Conclusion}
Agentic AI turns inference from batchable completions into sequential trajectories, moving the hardware bottleneck from throughput to single-stream decode latency-a quantity governed by weight-tier bandwidth, not peak compute. SRAM-resident accelerators exploit this directly, and by 2026 the industry, including the GPU incumbent, has converged on pairing them with HBM-rich prefill engines. Reliability compounds exponentially in trajectory length, so determinism and tail latency are performance variables. The evaluation question is therefore no longer ``GPU or specialised?'' but which prefill/decode pair, through which channel, at what utilisation-measured as goodput at an agentic SLO and cost per successful episode, with paired statistics, attested vendor evidence, and break-even conditions on capital ratio, utilisation and adoption lead time. The framework is built so that its placeholders, once populated, yield a reproducible cross-platform study; H7 is the result worth running it for.

\bibliography{main}

@inproceedings{vaswani2017,
  author      = {Ashish Vaswani and Noam Shazeer and Niki Parmar and {et al.}},
  title       = {Attention is All You Need},
  booktitle   = {Advances in Neural Information Processing Systems (NeurIPS) 30},
  year        = {2017}
}

@inproceedings{pope2023,
  author      = {Reiner Pope and Sholto Douglas and Aakanksha Chowdhery and {et al.}},
  title       = {Efficiently Scaling Transformer Inference},
  booktitle   = {Proceedings of Machine Learning and Systems (MLSys)},
  year        = {2023}
}

@article{williams2009,
  author      = {Samuel Williams and Andrew Waterman and David Patterson},
  title       = {Roofline: An Insightful Visual Performance Model for Multicore Architectures},
  journal     = {Communications of the ACM},
  number      = {4},
  volume      = {52},
  pages       = {65--76},
  year        = {2009}
}

@article{wulf1995,
  author      = {Wm. A. Wulf and Sally A. McKee},
  title       = {Hitting the Memory Wall: Implications of the Obvious},
  journal     = {ACM SIGARCH Computer Architecture News},
  number      = {1},
  volume      = {23},
  pages       = {20--24},
  year        = {1995}
}

@article{gholami2024,
  author      = {Amir Gholami and Zhewei Yao and Sehoon Kim and {et al.}},
  title       = {{AI} and Memory Wall},
  journal     = {IEEE Micro},
  number      = {3},
  volume      = {44},
  pages       = {33--39},
  year        = {2024}
}

@inproceedings{kwon2023,
  author      = {Woosuk Kwon and Zhuohan Li and Siyuan Zhuang and {et al.}},
  title       = {Efficient Memory Management for Large Language Model Serving with {PagedAttention}},
  booktitle   = {Proceedings of the 29th ACM Symposium on Operating Systems Principles (SOSP)},
  pages       = {611--626},
  year        = {2023}
}

@inproceedings{yu2022orca,
  author      = {Gyeong-In Yu and Joo Seong Jeong and Geon-Woo Kim and {et al.}},
  title       = {Orca: A Distributed Serving System for Transformer-Based Generative Models},
  booktitle   = {Proceedings of the 16th USENIX Symposium on Operating Systems Design and Implementation (OSDI)},
  pages       = {521--538},
  year        = {2022}
}

@inproceedings{agrawal2024sarathi,
  author      = {Amey Agrawal and Nitin Kedia and Ashish Panwar and {et al.}},
  title       = {Taming Throughput-Latency Tradeoff in {LLM} Inference with {Sarathi-Serve}},
  booktitle   = {Proceedings of the 18th USENIX Symposium on Operating Systems Design and Implementation (OSDI)},
  pages       = {117--134},
  year        = {2024}
}

@inproceedings{zhong2024distserve,
  author      = {Yinmin Zhong and Shengyu Liu and Junda Chen and {et al.}},
  title       = {{DistServe}: Disaggregating Prefill and Decoding for Goodput-optimized Large Language Model Serving},
  booktitle   = {Proceedings of the 18th USENIX Symposium on Operating Systems Design and Implementation (OSDI)},
  pages       = {193--210},
  year        = {2024}
}

@inproceedings{patel2024splitwise,
  author      = {Pratyush Patel and Esha Choukse and Chaojie Zhang and {et al.}},
  title       = {Splitwise: Efficient Generative {LLM} Inference Using Phase Splitting},
  booktitle   = {Proceedings of the 51st ACM/IEEE International Symposium on Computer Architecture (ISCA)},
  pages       = {118--132},
  year        = {2024}
}

@inproceedings{leviathan2023,
  author      = {Yaniv Leviathan and Matan Kalman and Yossi Matias},
  title       = {Fast Inference from Transformers via Speculative Decoding},
  booktitle   = {Proceedings of the 40th International Conference on Machine Learning (ICML)},
  pages       = {19274--19286},
  year        = {2023}
}

@inproceedings{ainslie2023gqa,
  author      = {Joshua Ainslie and James Lee-Thorp and Michiel de Jong and {et al.}},
  title       = {{GQA}: Training Generalized Multi-Query Transformer Models from Multi-Head Checkpoints},
  booktitle   = {Proceedings of the Conference on Empirical Methods in Natural Language Processing (EMNLP)},
  pages       = {4895--4901},
  year        = {2023}
}

@INPROCEEDINGS{Abbas2024,
  author={A. R. Ali and K. Kumar and M. A. Siddiqui and M. Zahid},
  booktitle={2024 International Joint Conference on Neural Networks (IJCNN)}, 
  title={An Open-source Cross-Industry and Cloud-agnostic Generative AI Platform}, 
  year={2024},
  volume={},
  number={},
  pages={1-10},
  doi={10.1109/IJCNN60899.2024.10650688}
}

@inproceedings{dao2022flash,
  author      = {Tri Dao and Daniel Y. Fu and Stefano Ermon and {et al.}},
  title       = {{FlashAttention}: Fast and Memory-Efficient Exact Attention with {IO}-Awareness},
  booktitle   = {Advances in Neural Information Processing Systems (NeurIPS) 35},
  year        = {2022}
}

@inproceedings{dao2024flash2,
  author      = {Tri Dao},
  title       = {{FlashAttention-2}: Faster Attention with Better Parallelism and Work Partitioning},
  booktitle   = {Proceedings of the International Conference on Learning Representations (ICLR)},
  year        = {2024}
}

@article{little1961,
  author      = {John D. C. Little},
  title       = {A Proof for the Queuing Formula: {$L=\lambda W$}},
  journal     = {Operations Research},
  number      = {3},
  volume      = {9},
  pages       = {383--387},
  year        = {1961}
}

@inproceedings{yao2023react,
  author      = {Shunyu Yao and Jeffrey Zhao and Dian Yu and {et al.}},
  title       = {{ReAct}: Synergizing Reasoning and Acting in Language Models},
  booktitle   = {Proceedings of the International Conference on Learning Representations (ICLR)},
  year        = {2023}
}

@article{snell2024,
  author      = {Charlie Snell and Jaehoon Lee and Kelvin Xu and {et al.}},
  title       = {Scaling {LLM} Test-Time Compute Optimally Can Be More Effective than Scaling Model Parameters},
  journal     = {arXiv preprint arXiv:2408.03314},
  year        = {2024}
}

@article{deepseek2025,
  author      = {{DeepSeek-AI}},
  title       = {{DeepSeek-R1} Incentivizes Reasoning in {LLMs} through Reinforcement Learning},
  journal     = {Nature},
  volume      = {645},
  pages       = {633--638},
  year        = {2025}
}

@techreport{openai2024o1,
  author      = {{OpenAI}},
  title       = {{OpenAI} o1 System Card},
  institution = {OpenAI},
  year        = {2024}
}

@article{wang2024survey,
  author      = {Lei Wang and Chen Ma and Xueyang Feng and {et al.}},
  title       = {A Survey on Large Language Model Based Autonomous Agents},
  journal     = {Frontiers of Computer Science},
  volume      = {18},
  pages       = {186345},
  year        = {2024}
}

@article{wu2023autogen,
  author      = {Qingyun Wu and Gagan Bansal and Jieyu Zhang and {et al.}},
  title       = {{AutoGen}: Enabling Next-Gen {LLM} Applications via Multi-Agent Conversation},
  journal     = {arXiv preprint arXiv:2308.08155},
  booktitle   = {arXiv preprint arXiv:2308.08155},
  year        = {2023}
}

@article{cemri2025,
  author      = {Mert Cemri and Melissa Z. Pan and Shuyi Yang and {et al.}},
  title       = {Why Do Multi-Agent {LLM} Systems Fail?},
  journal     = {arXiv preprint arXiv:2503.13657},
  year        = {2025}
}

@article{kapoor2024,
  author      = {Sayash Kapoor and Benedikt Stroebl and Zachary S. Siegel and {et al.}},
  title       = {{AI} Agents That Matter},
  journal     = {arXiv preprint arXiv:2407.01502},
  year        = {2024}
}

@article{yao2024tau,
  author      = {Shunyu Yao and Noah Shinn and Pedram Razavi and {et al.}},
  title       = {{$\tau$-bench}: A Benchmark for Tool-Agent-User Interaction in Real-World Domains},
  journal     = {arXiv preprint arXiv:2406.12045},
  year        = {2024}
}

@inproceedings{jimenez2024swebench,
  author      = {Carlos E. Jimenez and John Yang and Alexander Wettig and {et al.}},
  title       = {{SWE-bench}: Can Language Models Resolve Real-World {GitHub} Issues?},
  booktitle   = {Proceedings of the International Conference on Learning Representations (ICLR)},
  year        = {2024}
}

@article{liu2024lost,
  author      = {Nelson F. Liu and Kevin Lin and John Hewitt and {et al.}},
  title       = {Lost in the Middle: How Language Models Use Long Contexts},
  journal     = {Transactions of the Association for Computational Linguistics},
  volume      = {12},
  pages       = {157--173},
  year        = {2024}
}

@inproceedings{reddi2020mlperf,
  author      = {Vijay Janapa Reddi and Christine Cheng and David Kanter and {et al.}},
  title       = {{MLPerf} Inference Benchmark},
  booktitle   = {Proceedings of the 47th ACM/IEEE International Symposium on Computer Architecture (ISCA)},
  pages       = {446--459},
  year        = {2020}
}

@misc{mlcommons2026,
  author      = {{MLCommons}},
  title       = {{MLPerf} Inference v5.0--v6.0 Results and Rules, Including the Interactive Scenario for {Llama 2 70B}},
  howpublished= {\url{https://mlcommons.org}},
  year        = {2026}
}

@misc{artificialanalysis,
  author      = {{Artificial Analysis}},
  title       = {Independent Benchmarks of {LLM} {API} Providers: Output Speed, Time-to-First-Token and Price},
  howpublished= {\url{https://artificialanalysis.ai}},
  year        = {2026}
}

@article{lie2023,
  author      = {Sean Lie},
  title       = {Cerebras Architecture Deep Dive: First Look Inside the Hardware/Software Co-Design for Deep Learning},
  journal     = {IEEE Micro},
  number      = {3},
  volume      = {43},
  pages       = {18--30},
  year        = {2023}
}

@misc{cerebras2024,
  author      = {{Cerebras Systems}},
  title       = {{WSE-3} and {CS-3} Announcement; {Cerebras} Inference Launch; {Llama 3.1 405B} at 969 Tokens per Second},
  howpublished= {Press releases, March, August and November 2024. \url{https://www.cerebras.ai}},
  year        = {2024}
}

@misc{cerebras2026s1,
  author      = {{Cerebras Systems}},
  title       = {Form {S-1} Registration Statement and First-Quarter 2026 Results},
  howpublished= {US Securities and Exchange Commission filing and investor release. \url{https://investors.cerebras.ai}},
  year        = {2026}
}

@misc{cerebras2026cs4,
  author      = {{Cerebras Systems}},
  title       = {Cerebras Unveils {CS-4}: Up to 30 Times Faster than {GPU}-Based Solutions},
  howpublished= {Press release, 18 August 2026. \url{https://investors.cerebras.ai}},
  year        = {2026}
}

@inproceedings{prabhakar2024sn40l,
  author      = {Raghu Prabhakar and Ram Sivaramakrishnan and Darshan Gandhi and {et al.}},
  title       = {{SambaNova SN40L}: Scaling the {AI} Memory Wall with Dataflow and Composition of Experts},
  booktitle   = {Proceedings of the 57th IEEE/ACM International Symposium on Microarchitecture (MICRO)},
  year        = {2024}
}

@misc{sambanova2026,
  author      = {{SambaNova Systems}},
  title       = {{SambaNova} Announcements: {DeepSeek-R1} at 198 Tokens per Second (2025); {SN50}, Intel Collaboration and Series {E} (February 2026); Series {F} and {JPMorganChase} Partnership (July 2026)},
  howpublished= {Press releases. \url{https://sambanova.ai}},
  year        = {2026}
}

@inproceedings{abts2020tsp,
  author      = {Dennis Abts and Jonathan Ross and Jonathan Sparling and {et al.}},
  title       = {Think Fast: A Tensor Streaming Processor ({TSP}) for Accelerating Deep Learning Workloads},
  booktitle   = {Proceedings of the 47th ACM/IEEE International Symposium on Computer Architecture (ISCA)},
  pages       = {145--158},
  year        = {2020}
}

@inproceedings{abts2022,
  author      = {Dennis Abts and Garrin Kimmell and Andrew Ling and {et al.}},
  title       = {A Software-Defined Tensor Streaming Multiprocessor for Large-Scale Machine Learning},
  booktitle   = {Proceedings of the 49th ACM/IEEE International Symposium on Computer Architecture (ISCA)},
  pages       = {567--580},
  year        = {2022}
}

@misc{nvidia2026,
  author      = {{NVIDIA}},
  title       = {{H100} and {H200} Datasheets; {Blackwell} Architecture Technical Brief; {GTC} 2026 Keynote: {Vera Rubin} Platform and {Groq 3 LPX}},
  howpublished= {Technical documentation and keynote, 2024--2026. \url{https://www.nvidia.com}},
  year        = {2026}
}

@misc{nvidia2026nvl72,
  author      = {{NVIDIA}},
  title       = {{GB200 NVL72} and {GB300 NVL72} Rack-Scale Platform Specifications},
  howpublished= {\url{https://www.nvidia.com/en-us/data-center/gb200-nvl72/}},
  year        = {2026}
}

@misc{groq2025,
  author      = {{Groq}},
  title       = {Statement on Non-Exclusive Licensing Agreement with {NVIDIA}},
  howpublished= {Company statement, 24 December 2025. \url{https://groq.com}},
  year        = {2025}
}

@misc{groq2024aa,
  author      = {{Groq}},
  title       = {{Llama 3 Instruct 8B/70B} on the {LPU} Inference Engine: {Artificial Analysis} Results; {Llama 3.3 70B} Model Documentation},
  howpublished= {Blog and documentation, 2024--2026. \url{https://groq.com}},
  year        = {2024}
}

@misc{intel2025,
  author      = {{Intel}},
  title       = {{Gaudi 3} {AI} Accelerator White Paper; {Crescent Island} Data-Centre {GPU} Announcement, {OCP} Global Summit},
  howpublished= {Technical documentation and announcement, 2024--2025. \url{https://www.intel.com}},
  year        = {2025}
}

@misc{aws2026,
  author      = {{Amazon Web Services} and {Cerebras Systems}},
  title       = {Disaggregated Inference Combining {Trainium}-based Prefill and {CS-3} Decode on {Amazon Bedrock}},
  howpublished= {Announcement, March 2026. \url{https://aws.amazon.com}},
  year        = {2026}
}

@inproceedings{jouppi2023tpuv4,
  author      = {Norman P. Jouppi and George Kurian and Sheng Li and {et al.}},
  title       = {{TPU} v4: An Optically Reconfigurable Supercomputer for Machine Learning with Hardware Support for Embeddings},
  booktitle   = {Proceedings of the 50th ACM/IEEE International Symposium on Computer Architecture (ISCA)},
  year        = {2023}
}

@article{hennessy2019,
  author      = {John L. Hennessy and David A. Patterson},
  title       = {A New Golden Age for Computer Architecture},
  journal     = {Communications of the ACM},
  number      = {2},
  volume      = {62},
  pages       = {48--60},
  year        = {2019}
}

@article{hooker2021,
  author      = {Sara Hooker},
  title       = {The Hardware Lottery},
  journal     = {Communications of the ACM},
  number      = {12},
  volume      = {64},
  pages       = {58--65},
  year        = {2021}
}

@book{barroso2018,
  author      = {Luiz Andr{\'e} Barroso and Urs H{\"o}lzle and Parthasarathy Ranganathan},
  title       = {The Datacenter as a Computer: Designing Warehouse-Scale Machines},
  publisher   = {Morgan \& Claypool},
  edition     = {3rd},
  year        = {2018}
}

@inproceedings{luccioni2024,
  author      = {Sasha Luccioni and Yacine Jernite and Emma Strubell},
  title       = {Power Hungry Processing: Watts Driving the Cost of {AI} Deployment?},
  booktitle   = {Proceedings of the ACM Conference on Fairness, Accountability, and Transparency (FAccT)},
  pages       = {85--99},
  year        = {2024}
}

@inproceedings{samsi2023,
  author      = {Siddharth Samsi and Dan Zhao and Joseph McDonald and {et al.}},
  title       = {From Words to Watts: Benchmarking the Energy Costs of Large Language Model Inference},
  booktitle   = {Proceedings of the IEEE High Performance Extreme Computing Conference (HPEC)},
  year        = {2023}
}

@techreport{iea2025,
  author      = {{International Energy Agency}},
  title       = {Energy and {AI}},
  institution = {IEA},
  address     = {Paris},
  year        = {2025}
}

@misc{dora2022,
  author      = {{European Parliament and Council}},
  title       = {Regulation ({EU}) 2022/2554 on Digital Operational Resilience for the Financial Sector ({DORA})},
  howpublished= {Official Journal of the European Union, L 333, 1--79},
  year        = {2022}
}

@misc{eba2019,
  author      = {{European Banking Authority}},
  title       = {Guidelines on Outsourcing Arrangements, {EBA/GL/2019/02}},
  year        = {2019}
}

@misc{pra2021,
  author      = {{Prudential Regulation Authority}},
  title       = {Supervisory Statement {SS2/21}: Outsourcing and Third Party Risk Management},
  howpublished= {Bank of England},
  year        = {2021}
}

@techreport{nist2024genai,
  author      = {{National Institute of Standards and Technology}},
  title       = {Artificial Intelligence Risk Management Framework: Generative Artificial Intelligence Profile},
  institution = {NIST},
  number      = {NIST AI 600-1},
  year        = {2024}
}

@misc{sr117,
  author      = {{Board of Governors of the Federal Reserve System} and {Office of the Comptroller of the Currency}},
  title       = {Supervisory Guidance on Model Risk Management, {SR 11-7} / {OCC Bulletin 2011-12}},
  year        = {2011}
}

@article{fedus2022,
  author      = {William Fedus and Barret Zoph and Noam Shazeer},
  title       = {Switch Transformers: Scaling to Trillion Parameter Models with Simple and Efficient Sparsity},
  journal     = {Journal of Machine Learning Research},
  volume      = {23},
  pages       = {1--39},
  year        = {2022}
}

@article{xu2024agentcompany,
  author      = {Frank F. Xu and Yufan Song and Boxuan Li and Yuxuan Tang and Kritanjali Jain and Mengxue Bao and Zora Z. Wang and Xuhui Zhou and Zhitong Guo and Murong Cao and Mingyang Yang and Hao Yang Lu and Amaad Martin and Zhe Su and Leander Maben and Raj Mehta and Wayne Chi and Lawrence Jang and Yiqing Xie and Shuyan Zhou and Graham Neubig},
  title       = {{TheAgentCompany}: Benchmarking {LLM} Agents on Consequential Real World Tasks},
  journal     = {arXiv preprint arXiv:2412.14161},
  year        = {2024}
}

@article{yehudai2025,
  author      = {Asaf Yehudai and Lilach Eden and Alan Li and Guy Uziel and Yilun Zhao and Roy Bar-Haim and Arman Cohan and Michal Shmueli-Scheuer},
  title       = {A Survey on Evaluation of {LLM}-based Agents},
  journal     = {arXiv preprint arXiv:2503.16416},
  year        = {2025}
}

@book{efron1993,
  author      = {Bradley Efron and Robert J. Tibshirani},
  title       = {An Introduction to the Bootstrap},
  publisher   = {Chapman \& Hall/CRC},
  year        = {1993}
}

@book{montgomery2019,
  author      = {Douglas C. Montgomery},
  title       = {Design and Analysis of Experiments},
  publisher   = {Wiley},
  edition     = {10th},
  year        = {2019}
}
\bibliographystyle{mlsys2025}

\appendix 











\appendix



\section*{Supplementary Appendix}
\label{app:appendix}

\subsection*{A. Facility Characteristics}\label{app:facility}
Table~\ref{tab:facility} summarises power and cooling characteristics. Two observations matter more than the figures: liquid cooling no longer differentiates wafer-scale from Blackwell- or Rubin-class GPU racks, and rack-level power densities converge, so the discriminating quantity is tokens per kilowatt-hour under the agentic workload, which only measurement can supply.
\begin{table}[H]\centering\scriptsize
\caption{Facility characteristics (vendor-stated or widely reported).}\label{tab:facility}
\begin{tabular}{L{2.5cm}L{2.1cm}L{2.7cm}}\toprule
\textbf{System} & \textbf{Power} & \textbf{Cooling / footprint}\\\midrule
Cerebras CS-3 & $\approx$23-25\,kW, 15U & liquid; 2 per rack $\approx$50\,kW\\
Cerebras CS-4 rack & n/d (3 wafers) & liquid; rack-scale\\
DGX H100 / B200 & $\approx$10-14\,kW and above & air (H100) / liquid (B200)\\
GB200 / GB300 NVL72 & $\approx$120-140\,kW & liquid, mandatory\\
Groq\,3 LPX rack & n/d & liquid, dense\\
SambaRack 16$\times$SN40L & $\sim$10-15\,kW & air, high airflow\\
Gaudi 3 (8 OAM) & $\approx$8-10\,kW & air or liquid\\
\bottomrule\end{tabular}
\end{table}

\subsection*{B. Vendor Readiness and Security Questions}\label{app:vendor}
Vendor viability is a formal input to third-party-risk assessment. Table~\ref{tab:vendor} records the landscape at the time of writing (September 2026); sources are the regulatory filings and press releases cited in the main text.
\begin{table}[t]\centering\scriptsize
\caption{Vendor readiness matrix, September 2026.}\label{tab:vendor}
\begin{tabular}{L{1.5cm}L{6.4cm}}\toprule
\textbf{Vendor} & \textbf{Status; anchors and channels; access modes; residual risks}\\\midrule
Cerebras & \emph{Status:} public (Nasdaq: CBRS, May 2026); 2025 revenue US\$510\,M; CS-4 announced Aug 2026. \emph{Anchors:} OpenAI 750\,MW; AWS Bedrock decode tier; G42/MBZUAI concentration. \emph{Access:} appliance; Cerebras Cloud; Bedrock; HPE and Supermicro. \emph{Risks:} customer concentration; wafer yield; export licences; single-appliance availability.\\
NVIDIA (incl.\ Groq LPU) & \emph{Status:} incumbent; US\$20\,B non-exclusive Groq licence; Groq\,3 LPX ships Q3 2026. \emph{Anchors:} universal. \emph{Access:} Vera Rubin racks with LPX decode tier; Dynamo disaggregated serving. \emph{Risks:} licence-structure scrutiny; LPX capacity; premium pricing.\\
Groq (independent) & \emph{Status:} GroqCloud continues; core team departed. \emph{Anchors:} developer cloud. \emph{Access:} API only. \emph{Risks:} roadmap and talent for on-prem hardware.\\
SambaNova & \emph{Status:} private; US\$1\,B Series F at US\$11\,B (Jul 2026). \emph{Anchors:} JPMorganChase on-prem; SoftBank SN50; US DOE labs. \emph{Access:} SambaRack; managed co-location; SambaCloud; Xeon\,6 heterogeneous design. \emph{Risks:} SN50 execution; support scale; proprietary stack.\\
Intel & \emph{Status:} Gaudi 3 de-emphasised; Crescent Island sampling H2 2026. \emph{Anchors:} IBM Cloud, Dell. \emph{Access:} OEM. \emph{Risks:} roadmap credibility.\\
AMD & \emph{Status:} MI355X shipping; leads MLPerf v6.0 Llama-2-70B rows. \emph{Anchors:} hyperscalers; OpenAI (reported). \emph{Access:} OEM; ROCm. \emph{Risks:} ecosystem maturity.\\
\bottomrule\end{tabular}
\end{table}
\paragraph{Security-architecture questions specific to inference appliances.}
(i) Management and telemetry planes: outbound connectivity, air-gapped operation and offline update cadence. (ii) Remote support: access path, authentication, session recording, and whether break-glass access can be disabled. (iii) Identity and isolation: enterprise identity integration and tenancy or memory isolation on a single wafer or dataflow fabric, which GPUs provide through hardware memory protection and virtualisation. (iv) Data residency in disaggregated mode: the KV cache in transit between prefill and decode pools is customer data; its encryption, residency and retention must be specified. (v) Supply chain: fabrication location, firmware provenance and export-licence exposure.

\subsection*{C. Internal Measurement Templates}\label{app:templates}
Result templates keyed to protocol axes A1, A3 and A5 are provided as CSV skeletons in the artifact repository: context $\times$ TTFT/TPOT P50/P99 and KV footprint per system; concurrency $\times$ goodput at $\theta_2=10/40$\,ms, P99 TPOT, wall power and tok/J; and $K\times\tau_{\mathrm{tool}}$ $\times$ $T_{\mathrm{ep}}$ P50/P95, decode idle fraction, prefix-cache hit rate, KV-transfer P99 and success $S$. No illustrative numbers are given, to avoid their being mistaken for data.


\end{document}